\documentclass[conference]{IEEEtran}

\usepackage[utf8]{inputenc}
\usepackage{fancyhdr}
\usepackage{amsmath}%
\usepackage{graphicx}

\usepackage{amsmath} 
\usepackage{amssymb} 
\usepackage{pifont} 
 \usepackage{cite}
\usepackage{url}
\usepackage{comment}
\usepackage{color, colortbl}
\usepackage{enumerate}
\usepackage{arcs}

\usepackage[framemethod=TikZ]{mdframed}

\usepackage{fancyhdr}

\usepackage{pdfpages}

\newtheorem{defini}{Definition}
\newtheorem{thm}{Theorem}

\newtheorem{exemple}{Example}

\definecolor{Gray}{gray}{0.9}
\begin{document}


\title{{On the Security of Cryptographic Protocols  Using the Little Theorem of Witness Functions}}




\author{\IEEEauthorblockN{Jaouhar Fattahi$^1$, Mohamed Mejri$^1$ and Emil Pricop$^2$}

\IEEEauthorblockA{$^1$Département d'informatique et de génie logiciel. Laboratoire de sécurité informatique (LSI) . Université Laval. \\\ Québec. Canada.\\}
\IEEEauthorblockA{$^2$Automatic Control, Computers and Electronics Department Petroleum-Gas University of Ploiesti\\  Ploiesti. Romania.\\}


}


%


\maketitle
\thispagestyle{plain}

\fancypagestyle{plain}{
\fancyhf{}	
\fancyfoot[L]{}
\fancyfoot[C]{}
\fancyfoot[R]{}
\renewcommand{\headrulewidth}{0pt}
\renewcommand{\footrulewidth}{0pt}
}

\pagestyle{fancy}{
\fancyhf{}
\fancyfoot[R]{}}
\renewcommand{\headrulewidth}{0pt}
\renewcommand{\footrulewidth}{0pt}

\begin{abstract}

In this paper, we show how practical the little theorem of witness functions is in detecting security flaws in some category of cryptographic protocols. We convey a formal analysis of the Needham-Schroeder symmetric-key protocol in the theory of witness functions. We show how it helps to teach about a security vulnerability in a given step of this protocol where the value of security of a particular sensitive ticket in a sent message unexpectedly plummets compared with its value when received. This vulnerability may be exploited by an intruder to mount a replay attack as described by Denning and Sacco. \\
                                                                                                                                                                                                                                                                                                                                                                                                                                                                                                                                                                                                                                                                                                                                                                                                                                                                                                                                                                                                                                                                                                                                                                                                                                                                                                                                                                                                                                                                                                                                                                                                                                                                                                                                                                                                                                                                                                                                                                                                                                                                                                                                                                                                                                                                                                                                                                                                                                                                                                                                                                                                                                                                                                                                                                                                                                                                                                                                                                                                                                                                                                                                                                                                                                                                                                                                                                                                                                                                                                                                                                                                                                                                                                                                                                                                                                                                                                                                                                                                                                                                                                                                                                                                                                                                                                                                                                                                                                                                                                                                                                                                                                                                                                                                                                                                                                                                                                                                                                                                                                                                                                                                                                                                                                                                                                                                                                                                                                                                                                                                                  
\end{abstract}


%
\IEEEpeerreviewmaketitle

\section*{Notice}
 ${\mbox{\scriptsize{\copyright}}}$ \textit{2019 IEEE. Personal use of this material is permitted. Permission from IEEE must be obtained for all other uses, in any current or future media, including reprinting/republishing this material for advertising or promotional purposes, creating new collective works, for resale or redistribution to servers or lists, or reuse of any copyrighted component of this work in other works.}

\section{Introduction}

Cryptographic protocols are distributed programs that aim to secure communications in an unsecured network by means of cryptography. They are used wherever the security matters, for example, in electronic commerce, military communications, electronic voting, etc. A security breach in a protocol often causes significant and irreparable damage. Indeed, a flaw in an online sales protocol can cause huge losses for a remote seller. The victim, on the other hand, may end up with a bad credit record and a tarnished reputation. The use of cryptography, while necessary, cannot guarantee the security of a protocol. Indeed, the history of protocols \cite{cortier:inria-00000552,10.1007/BFb0000430} teaches us that an intruder is capable of manipulating the rules of a given protocol to infer a secret, usurp an identity, alter the integrity of data or deny his participation in a given communication after committing an unhealthy act. \\

Formal methods have therefore emerged \cite{DBLP:conf/csfw/Schneider97,DBLP:journals/jacm/AbadiB05,DBLP:conf/csfw/Blanchet18,DBLP:journals/jancl/ArmandoCC09,DBLP:conf/fosad/Abadi07,DBLP:conf/csfw/ChevalCT18, DBLP:conf/cav/ChevalKR18} as a preferred means of verifying whether a protocol meets the security properties for which it was intended. As a result, a number of methods have sprung up, and have demonstrated distinct performance. Others were withdrawn after a period of glory due to unfortunate deficiencies, several years after their release.\\

In recent years, a new generation of analytical functions, called witness functions \cite{ChapterJFSpringer,TheseJF,8123025,DBLP:conf/apn/FattahiMH14}, has been proposed to analyze cryptographic protocols. These functions assign a security value to each message component exchanged in the protocol, then examine whether this value increases or not between two reception-sending steps. If throughout the protocol, all security values are observed increasing, then the protocol itself is said to be increasing and hence declared correct for the  property of secrecy. If, on the other hand, a decreasing value is found, these functions refuse to certify the correctness of the protocol. This way of approaching and handling security in cryptographic protocols stems from the fact that increasing protocols are correct, proven in \cite{TheseJF}. These functions have been a good way to demonstrate the correctness of several protocols. They were also able to teach about security vulnerabilities in other protocols. More recently, in \cite{8447762}, the author proposed a reduced form of the general theorem of protocol correctness by witness functions \cite{DBLP:conf/apn/FattahiMH14} for analyzing tagged protocols. The author adopts a very broad definition of tagged protocols and considers tagged any protocol whose messages are distinguishable, one by one by a receiving agent, either by inserting a particular syntactic element in a message, or thanks to the position of identities or nonces, or by any other means.\\

In this paper, we  provide an analysis of the Needham-Schroeder symmetric-key protocol \cite{Needham1978} using this reduced theorem that we refer to by the little theorem of witness functions as opposed to the general theorem. We  show that this protocol does not respect this theorem and gets stuck in a critical step in the protocol where a value of security unexpectedly goes down which may be interpreted as a security vulnerability. We also show that this latter can be exploited by an intruder using a Denning and Sacco attack scenario \cite{Denning}.

\section{Paper organization}

The paper is organized as follows:

\begin{itemize}
\item in Section III, we briefly present the role-based specification in which a protocol is specified;
 \item in Section IV, we shortly review the foundation of witness functions;
  \item in Section V, we present the little theorem of witness functions;
  \item in Section VI, we give a formal analysis of the Needham-Schroeder symmetric-key protocol using this theorem;
  \item in Section VII, we discuss the results of our analysis;
  \item in Section VIII, we compare our approach with other related approaches;
  \item in Section IX, we conclude.
\end{itemize}

\section{Role-based specification}

A cryptographic protocol is a set of programs that can communicate over the network. Each program corresponds to a role of the protocol. Agents, who are actually servers or people who can implement the protocol, can play several roles simultaneously. When an honest agent takes on a role, an instance of the program corresponding to the role runs on the agent's machine using its various personal data, for example its identity and secret keys. A session is an instance of a program executed by an agent. A dishonest agent, or intruder \cite{10.1007/978-3-540-24727-2_1}, is not required to follow a protocol role. It can execute any number of sessions. It has other capacities like intercepting messages, concatenating or de-concatenating messages, encrypting or decrypting messages with keys that it knows, etc.\\

A role-based specification \cite{DBLP:conf/acsac/DebbabiLM98, Mejri} is an abstraction of all these facts. It focuses on a single agent at a time and represents what, how, and to/from whom it sends and receives messages. If a component of a given message is not intelligible for that agent, it is replaced by a variable. In that case, we talk about generalized roles. An exponent is added to represent the session. 

\section{Witness Functions} 

\subsection{Reliable Function}\label{sectionFonctionsetSelections}

Let's consider the following function $F$:\\

\begin{defini}{(Reliable Function)}\label{reliable}
\scalebox{0.9}{
\begin{tabular}{lrcl}
 1.& ${F}(\alpha,\{\alpha\})$ & $=$ & $\bot$ \\
  2. & ${F}(\alpha, {M}_1 \cup {M}_2)$ & $=$ & ${F}(\alpha, {M}_1)\sqcap{F}(\alpha,{M}_2)$ \\
   3. & ${F}(\alpha,{M})$ & $=$ & $\top, \mbox{ if } \alpha \notin {\mathcal{A}}({M})$ \\
   4. & ${F}(\alpha,m_1.m_2)$ & $=$ & $F(\alpha,\{m_1,m_2\})$ \\
   5. & ${F}(\alpha,\{m\}_{k})$ & $=$ & $F(\alpha,\{m\}),$ \mbox{ if } $\ulcorner k^{-1} \urcorner \not \sqsupseteq \ulcorner \alpha \urcorner$\\
   6. & ${F}(\alpha,\{m\}_{k})$ & $=$ & $\ulcorner k^{-1} \urcorner  \cup \mbox{ID}(m),$ \mbox{ if } $\ulcorner k^{-1} \urcorner \sqsupseteq \ulcorner \alpha \urcorner$\\
\end{tabular}
}
\end{defini}
$ $\\
A reliable function is a function that assigns to each atom in a message a reasonable value of security. In the point 1., it assigns to a plain (clear) atomic message $\alpha$ the bottom value of security (i.e. $\bot$).  In the point 2., it assigns for an atomic message $\alpha$ that shows up in two sets of messages the minimum (i.e. $\sqcap$) of the two values calculated in each set separately. In the point 3., it assigns to an atomic message $\alpha$ that does not appear in a message (i.e. $\alpha \notin {\mathcal{A}}({M})$ where ${\mathcal{A}}({M})$ is the set of all atoms of $M$) the top value (i.e. $\top$). In the point 4., it assigns to an atomic message in a concatenated message the minimum of the values calculated in each message separately. In the point 5., $F$ disregards an encryption with an outer  key of a lower level than the analyzed atom (i.e. $\ulcorner k^{-1} \urcorner \not \sqsupseteq \ulcorner \alpha \urcorner$) and seeks a deeper strong key. In the point 6., $F$ first ensures that $\alpha$ is encrypted with a key $k$ of a higher level (i.e. $\ulcorner k^{-1} \urcorner \sqsupseteq \ulcorner \alpha \urcorner$) and returns the set of identities of agents that detain the reverse key (i.e. $\ulcorner k^{-1} \urcorner$) in addition to the set of identities in the neighborhood of $\alpha$ under the same encryption in $m$ (i.e. $\mbox{ID}(m)$). \\

\begin{exemple}\label{exemple1}
Say that we have a context such that: $\ulcorner \alpha \urcorner=\{A, B, S\}$; ${k_{ac}^{-1}}={k_{ac}},{k_{ab}^{-1}}={k_{ab}}, {k_{as}^{-1}}={k_{as}}$; $\ulcorner{k_{ac}}\urcorner=\{A, C\},\ulcorner{k_{as}}\urcorner=\{A, S\}, \ulcorner{k_{ab}}\urcorner=\{A, B\}$. Say that $m=\{\{C.\{E.\alpha.D\}_{k_{as}}\}_{k_{ab}}\}_{k_{ac}}$.

	\begin{tabular}{lcll}
		$F(\alpha,m)$&  =  &$F(\alpha,\{\{C.\{E.\alpha.D\}_{k_{as}}\}_{k_{ab}}\}_{k_{ac}})$& \\
		&  &\!\!\!\!\!\!\!\!\!\!\!\!\!\!\!\!\!\!\!\!\!\!\!\!\!\!\!\{Definition \ref{reliable}, Point 5., since $\ulcorner k_{ac}^{-1}\urcorner \not \sqsupseteq \ulcorner \alpha \urcorner$\}  &  \\
		&  =  &$F(\alpha,\{C.\{E.\alpha.D\}_{k_{as}}\}_{k_{ab}})$& \\
		&  &\!\!\!\!\!\!\!\!\!\!\!\!\!\!\!\!\!\!\!\!\!\!\!\!\!\!\!\{Definition \ref{reliable}, Point 6.\}  &  \\
		& = & $ \ulcorner{k_{ab}^{-1}}\urcorner { \cup} \{E, C, D\}   $ & \\
		&   &\!\!\!\!\!\!\!\!\!\!\!\!\!\!\!\!\!\!\!\!\!\!\!\!\!\!\!\{Since $\ulcorner k_{ab}^{-1}\urcorner = \{A, B\}$ in the context\}&  \\
		& = & $\{A, B, E, C, D\}$& \\
	\end{tabular}

\end{exemple} 
 $ $\\
In \cite{DBLP:conf/apn/FattahiMH14,TheseJF} we prove that $F$ is reliable and we define other reliable functions that we do not mention here. 

\subsection{Derivative Function}\label{derivativeFunc}

The function $F$ given in Definition \ref{reliable} does not deal with variables. In order to do so, we rather use its derivative form $F'$. This derivative function simply removes all variables around any evaluated atom before applying $F$. The following example explains how to use $F'$ to assign values of security to atomic messages in messages containing variables.\\

\begin{exemple}\label{exemple2}
Let us have the same context as Example \ref{exemple1}. Say that $m=\{\{C.\{X.\alpha.D\}_{k_{as}}\}_{k_{ab}}\}_{k_{ac}}$ where $X$ is a variable.

	\begin{tabular}{lcll}
		$F'(\alpha,m)$&  =  &$F'(\alpha,\{\{C.\{X.\alpha.D\}_{k_{as}}\}_{k_{ab}}\}_{k_{ac}})$& \\
		&  &\!\!\!\!\!\!\!\!\!\!\!\!\!\!\!\!\!\!\!\!\!\!\!\!\!\!\!\!\!\!\!\!\!\!\!\!\!\!\!\!\!\!\!\!\!\!\!\!\!\!\!\!\!\!\{The variable X is first removed by derivation, then $F$ is applied\}  &  \\
		&  =  &$F(\alpha,\{\{C.\{\alpha.D\}_{k_{as}}\}_{k_{ab}}\}_{k_{ac}})$& \\
		&  &\!\!\!\!\!\!\!\!\!\!\!\!\!\!\!\!\!\!\!\!\!\!\!\!\!\{Definition \ref{reliable}, Point 5., since $\ulcorner k_{ac}^{-1}\urcorner \not \sqsupseteq \ulcorner \alpha \urcorner$\}  &  \\
		&  =  &$F(\alpha,\{C.\{\alpha.D\}_{k_{as}}\}_{k_{ab}})$& \\
		&  &\!\!\!\!\!\!\!\!\!\!\!\!\!\!\!\!\!\!\!\!\!\!\!\!\!\{Definition \ref{reliable}, Point 6.\}  &  \\
		& = & $ \ulcorner{k_{ab}^{-1}}\urcorner { \cup} \{C, D\}   $ & \\
		&   &\!\!\!\!\!\!\!\!\!\!\!\!\!\!\!\!\!\!\!\!\!\!\!\!\!\{Since $\ulcorner k_{ab}^{-1}\urcorner = \{A, B\}$ in the context\}&  \\
		& = & $\{A, B, C, D\}$& \\
	\end{tabular}
\end{exemple} 
 $ $\\
The derivative function $F'$ is good for analyzing generalized roles (involving messages with variables) in the case of tagged protocols. However, it may present some complexities in the general case, which is beyond the scope of this paper. In the context of tagged protocols, $F'$ is called witness function. We recall that tagged protocols in our own definition are all protocols that generate messages such that they are one by one distinguishable from a receiver point of view, meaning, no regular message can be unified with a non regular one.
  
\section{Little Theorem of Witness Functions} \label{sec3}

\begin{thm}{[Little Theorem of Witness Functions]}\label{PATTag}
Let $p$ be a tagged protocol. Let $F'$ be a witness function.
$p$ is correct for secrecy if:
$\forall R.r \in R_G(p), \forall \alpha \in {\mathcal{A}}{(r^+ )}$ we have:
$$F'(\alpha,r^+) \sqsupseteq \ulcorner \alpha \urcorner \sqcap F'(\alpha, R^-)$$
\end{thm}

Theorem \ref{PATTag}, which we call the little theorem of the witness functions, uses the witness function $F'$ to attribute a value of security to every component in every single exchanged message in the protocol, then, it examines if the value of security of that component is increasing in the sent message (i.e. in $r^+$) compared with its value in either the context or when it was received (i.e. in $R^-$). If this is the case in all the protocol steps, then the protocol is declared correct for secrecy. Otherwise, the theorem refuses to certify the protocol correctness and flags up a possible vulnerability that could be exploited by an intruder. We will refer to this theorem by LTWF in the following analysis of the Needham-Schroeder symmetric key protocol.


\normalsize
\section{Formal analysis of the Needham-Schroeder symmetric-key protocol}\label{sec4}
In this section,  we analyze the Needham-Schroeder symmetric-key protocol with Theorem \ref{PATTag} (LTWF) for secrecy. This protocol is described in Table  \ref{NSLVar} and we refer to by $p$.
\begin{table}
\begin{center}
\begin{tabular}{|llllll|}
\hline
 $p$= &$\langle1,$ & $A$ & $\longrightarrow$ & $S:$ & $A.B.N_a\rangle$ \\
  &$\langle2,$ & $S$ & $\longrightarrow$ & $A:$ & $\{N_a.k_{ab}.B.\{k_{ab}.A\}_{k_{bs}}\}_{k_{as}}\rangle$ \\
  &$\langle3,$ & $A$ & $\longrightarrow$ & $B:$ & $\{k_{ab}.A\}_{k_{bs}}\rangle$\\
&$\langle4,$ & $B$ &$\longrightarrow$ & $A:$ & $\{N_b\}_{k_{ab}}\rangle$\\
&$\langle5,$ & $A$ &$\longrightarrow$ & $B:$ & $\{N_b - 1\}_{k_{ab}}\rangle$\\
\hline
\end{tabular}
\end{center}
\caption{The Needham-Schroeder symmetric-key protocol}
\label{NSLVar}
\end{table}

\subsection{Context setting}

The generalized roles of the protocol are  $\cal{R}_{\cal{G}}(\textit{p})=\{ A_{\cal{G}},B_{\cal{G}}, S_{\cal{G}}\}$ where:

\begin{center}
\scalebox{0.95}{
\begin{tabular}{lllllll}

  $\cal{A}_{\cal{G}}=$ & $i.1$ & $A$& $\longrightarrow I(S)$&:&$A.B.N_a^i$ \\
  $$ & $i.2$ & $I(S)$&$ \longrightarrow A$ & : & $\{N_a^i.X.B.Y\}_{k_{as}}$ \\
  $$ & $i.3$ & $A$& $\longrightarrow I(B)$ & : & $Y$ \\
  $$ & $i.4$ & $I(B)$& $\longrightarrow A$ & : & $\{Z\}_{X}$ \\
  $$ & $i.5$ & $A$& $\longrightarrow I(B)$ & : & $\{Z-1\}_{X}$ \\ $ $\\
  $\cal{B}_{\cal{G}}=$ & $j.1$ & $I(A)$& $\longrightarrow B$ & : & $\{T.A\}_{k_{bs}}  $\\
  $$ & $j.2$ & $B$&$ \longrightarrow I(A)$ & : & $\{N_b^j\}_{T}$ \\
  $$ & $j.3$ & $I(A)$&$ \longrightarrow B$ & : & $\{N_b^j - 1\}_{T}$ \\ $ $\\
 $\cal{S}_{\cal{G}}=$ & $n.1$ & $I(A)$& $\longrightarrow S$ & : & $A.B.Q$\\
 & $n.2$ & $S$& $\longrightarrow I(A)$ & : & $\{Q.k_{ab}.B.\{k_{ab}.A\}_{k_{bs}}\}_{k_{as}}$
\end{tabular}
}
\end{center}

\normalsize
$ $\\
Context :\\ 
$\ulcorner A\urcorner= \bot$; $\ulcorner B\urcorner= \bot$; $\ulcorner S\urcorner= \bot$; (i.e. three public identities)\\
$\ulcorner N_a\urcorner= \bot$ (i.e. public nonce);\\ 
$\ulcorner N_b\urcorner= \{A,B\}$ (i.e. nonce shared between $A$ and $B$); \\
$\ulcorner k_{ab}\urcorner=\{A, B, S\}$; (i.e. session key shared between $A$ and $B$ and created by $S$)\\
$\ulcorner k_{as}\urcorner=\{A, S\}$; (i.e. shared key between $A$ and $S$)\\
$\ulcorner k_{bs}\urcorner=\{B, S\}$; (i.e. shared key between $B$ and $S$)\\
$({\cal{L}},\sqsupseteq, \sqcup, \sqcap, \bot,\top)=(2^{\cal{I}},\subseteq,\cap,\cup,\cal{I}, \emptyset)$; (i.e. security lattice)\\
${\cal{I}}=\{I, A, B, S\}$; (i.e. intruder and regular agents participating in the protocol)\\
${\cal{X}}_p=\{X, Y, Z, T, Q\}$ is the set of variables. $F'$ is the used witness function and the derivative function of $F$ given in Definition \ref{reliable}. \\

It is wise to notice that all the generated messages by the protocol cannot overlap one with another from a receiver point of view, which means that we can use Theorem LTWF to analyze the protocol.

\subsection{Analyzing the generalized role of  $A$}

From the generalized role $\cal{A}_{\cal{G}}$, an agent $A$ may participate in three  receiving/sending steps. In the first step, it receives nothing and sends the message $A.B.N_a^i$. In the second step, it receives the message $\{N_a^i.X.B.Y\}_{k_{as}}$ and sends the message $Y$. In the third step, it receives the message $\{Z\}_{X}$ and sends the message $\{Z-1\}_{X}$. This is represented by the following  rules.
\[{S_{A}^{1}}:\frac{\Box}{A.B.N_a^i}; ~~~~~~ {S_{A}^{2}}:\frac{\{N_a^i.X.B.Y\}_{k_{as}}}{Y}; ~~~~~~ {S_{A}^{3}}:\frac{\{Z\}_{X}}{\{Z-1\}_{X}} \]

\subsubsection{Analyzing exchanged messages in $S_{A}^{1}$}
$ $\\
$ $\\
1- For $N_a^i$:$ $\\
$ $\\
a- On sending: $r_{S_{A}^{1}}^+=A.B.N_a^i$ \\
 
\scalebox{0.99}{
\begin{tabular}{lcll}
$F'(N_a^i,r_{S_{A}^{1}}^+)$ & $=$ & $F'(N_a^i,A.B.N_a^i)$& \\
&  &\!\!\!\!\!\!\!\!\!\!\!\!\!\!\!\!\!\!\!\!\!\!\!\!\!\!\!\!\!\!\!\!\!\!\!\!\!\!\!\!\!\!\!\!\!\!\!\!\!\!\!\!\!\!\!\!\!\!\!\!\!\!\!\!\!\!\!\!\!\!\{No variable in the neighborhood of $N_a^i$ to be removed by derivation\}  &  \\
 & = & $F(N_a^i,A.B.N_a^i)$ & \\
 &   &\!\!\!\!\!\!\!\!\!\!\!\!\!\!\!\!\!\!\!\!\!\!\!\!\!\!\!\!\!\!\!\!\!\!\!\!\!\!\!\!\!\!\!\!\!\!\!\!\!\! \{Definition \ref{reliable} (no encryption)\}&  \\
 & = & $\bot$~~~~~~~~~~~~~~~~~~~~~~~~~~~~~~~~~~~(1.1)& \\
\end{tabular}
}
$ $\\
b- On receiving: $R_{S^{i}}^-=\emptyset$\\
$ $\\
\scalebox{0.99}{
\begin{tabular}{lcll}
$F'(N_a^i,R_{S_{A}^{1}}^-)$ & $=$ & $F'(N_a^i,\emptyset)$& \\
&  &\!\!\!\!\!\!\!\!\!\!\!\!\!\!\!\!\!\!\!\!\!\!\!\!\!\!\!\!\!\!\!\!\!\!\!\!\!\!\!\!\!\!\!\!\!\!\!\!\!\!\!\!\!\!\!\!\!\!\!\!\!\!\!\!\!\{No variable in the neighborhood of $N_a^i$ to be removed by derivation\}  &  \\
 & = & $F(N_a^i,\emptyset)$ & \\
 &   &\!\!\!\!\!\!\!\!\!\!\!\!\!\!\!\!\!\!\!\!\!\!\!\!\!\!\!\{Definition \ref{reliable}\}&  \\
 & = & $\top$~~~~~~~~~~~~~~~~~~~~~~~~~~~~~~~~~~~~~~(1.2)& \\
\end{tabular}
}
$ $\\
2- Concordance with Theorem LTWF:\\
$ $\\
From 1.2 and since $\ulcorner N_a\urcorner= \{A,B\}$, we have: \\
$ $\\
$\ulcorner N_a^i\urcorner \sqcap F'(N_a^i,R_{S_{A}^{1}}^-)  =\bot \sqcap \top =\bot$ ~~~~~~~~~~~~~~~~~~~~~~~~~~~~~~~~~~~~~(1.3)\\
$ $\\
From 1.1 and 1.3, we have : \\
$ $\\
$F'(N_a^i,r_{S_{A}^{1}}^+) \sqsupseteq \ulcorner N_a^i\urcorner \sqcap F'(N_a^i,R_{S_{A}^{1}}^-)$ ~~~~~~~~~~~~~~~~~~~~~~(1.4)\\
$ $\\
From  1.4, $S_{A}^{1}$ respects Theorem LTWF.  ~~~~~~~~~~~~~~~~~~~~~(I)\\

\subsubsection{Analyzing exchanged messages in $S_{A}^{2}$}
$ $\\$ $\\
1- For $N_a^i$:$ $\\
$ $\\
a- On sending: $r_{S_{A}^{2}}^+=Y$ \\
$ $\\
\scalebox{0.99}{
\begin{tabular}{lcll}
$F'(N_a^i,r_{S_{A}^{2}}^+)$ & $=$ & $F'(N_a^i,Y)$& \\
&  &\!\!\!\!\!\!\!\!\!\!\!\!\!\!\!\!\!\!\!\!\!\!\!\!\!\!\!\!\!\!\!\!\!\!\!\!\!\!\!\!\!\!\!\!\!\!\!\!\!\! \{The variable $Y$ is removed by derivation\}  &  \\
 & = & $F(N_a^i,\emptyset)$ & \\
 &   &\!\!\!\!\!\!\!\!\!\!\!\!\!\!\!\!\!\!\!\!\!\!\!\{Definition \ref{reliable}\}&  \\
 & = & $\top$~~~~~~~~~~~~~~~~~~~~~~~~~~~~~~~~~~~~~~(2.1)& \\
\end{tabular}
}
$ $\\
b- On receiving: $R_{S_{A}^{2}}^-=\{N_a^i.X.B.Y\}_{k_{as}}$\\
$ $\\
\scalebox{0.99}{
\begin{tabular}{lcll}
$F'(N_a^i,R_{S_{A}^{2}}^-)$ & $=$ & $F'(N_a^i,\{N_a^i.X.B.Y\}_{k_{as}})$& \\
&  &\!\!\!\!\!\!\!\!\!\!\!\!\!\!\!\!\!\!\!\!\!\!\!\!\!\!\!\!\!\!\!\!\!\!\!\!\!\!\!\!\!\!\!\!\!\!\!\!\!\! \{The variables $X$ and $Y$ are removed by derivation\}  &  \\
 & = & $F(N_a^i,\{N_a^i.B\}_{k_{as}})$ & \\
 &   &\!\!\!\!\!\!\!\!\!\!\!\!\!\!\!\!\!\!\!\!\!\!\!\!\!\!\!\! \{Definition \ref{reliable} and since $\ulcorner k_{as}^{-1}\urcorner$ = \{A, S\}\}&  \\
 & = & $\{A, S, B\}$~~~~~~~~~~~~~~~~~~~~~~~~(2.2)& \\
\end{tabular}
}
$ $\\
2- For $X$:$ $\\
$ $\\
c- On sending: $r_{S_{A}^{2}}^+=Y$ \\
$ $\\
\scalebox{0.99}{
\begin{tabular}{lcll}
$F'(X,r_{S_{A}^{2}}^+)$ & $=$ & $F'(X,Y)$& \\
&  &\!\!\!\!\!\!\!\!\!\!\!\!\!\!\!\!\!\!\!\!\!\!\!\!\!\!\!\!\!\!\!\!\!\!\!\!\!\!\!\!\!\!\!\!\!\!\!\!\!\!\{The variable $Y$ is removed by derivation\}  &  \\
 & = & $F(X,\emptyset)$ & \\
 &   &\!\!\!\!\!\!\!\!\!\!\!\!\!\!\!\!\!\!\!\!\!\!\!\!\!\!\!\!  \{Definition \ref{reliable}\}&  \\
 & = & $\top$~~~~~~~~~~~~~~~~~~~~~~~~~~~~~~~~~~~(2.3)& \\
\end{tabular}
}
$ $\\
d- On receiving: $R_{S_{A}^{2}}^-=\{N_a^i.X.B.Y\}_{k_{as}}$\\
$ $\\
\scalebox{0.99}{
\begin{tabular}{lcll}
$F'(X,R_{S_{A}^{2}}^-)$ & $=$ & $F'(X,\{N_a^i.X.B.Y\}_{k_{as}})$& \\
&  &\!\!\!\!\!\!\!\!\!\!\!\!\!\!\!\!\!\!\!\!\!\!\!\!\!\!\!\!\!\!\!\!\!\!\!\!\!\!\!\!\!\!\!\!\!\!\!\!\!\!\!\!\!\!\{The variable $Y$ is removed by derivation\}  &  \\
 & = & $F(X,\{N_a^i.X.B\}_{k_{as}})$ & \\
 &   &\!\!\!\!\!\!\!\!\!\!\!\!\!\!\!\!\!\!\!\!\!\!\!\!\!\!\!\!\!\!\!\!\!\!\!\!\{Definition \ref{reliable} and since $\ulcorner k_{as}^{-1}\urcorner$ = \{A, S\}\}&  \\
 & = & $\{A, S, B\}$~~~~~~~~~~~~~~~~~~~~~~~~~(2.4)& \\
\end{tabular}
}
$ $\\
3- For $Y$:$ $\\
$ $\\
e- On sending: $r_{S_{A}^{2}}^+=Y$ \\
$ $\\
\scalebox{0.99}{
	\begin{tabular}{lcll}
		$F'(Y,r_{S_{A}^{2}}^+)$ & $=$ & $F'(Y,Y)$& \\
		&   &\!\!\!\!\!\!\!\!\!\!\!\!\!\!\!\!\!\!\!\!\!\!\!\!\!\!\!\!  \{Definition \ref{reliable}\}&  \\
		& = & $\bot$~~~~~~~~~~~~~~~~~~~~~~~~~~~~~~~~~~~(2.5)& \\
	\end{tabular}
}
$ $\\
f- On receiving: $R_{S_{A}^{2}}^-=\{N_a^i.X.B.Y\}_{k_{as}}$\\
$ $\\
\scalebox{0.99}{
	\begin{tabular}{lcll}
		$F'(Y,R_{S_{A}^{2}}^-)$ & $=$ & $F'(Y,\{N_a^i.X.B.Y\}_{k_{as}})$& \\
		&  &\!\!\!\!\!\!\!\!\!\!\!\!\!\!\!\!\!\!\!\!\!\!\!\!\!\!\!\!\!\!\!\!\!\!\!\!\!\!\!\!\!\!\!\!\!\!\!\!\!\!\!\!\!\!\{The variable $X$ is removed by derivation\}  &  \\
		& = & $F(Y,\{N_a^i.B.Y\}_{k_{as}})$ & \\
		&   &\!\!\!\!\!\!\!\!\!\!\!\!\!\!\!\!\!\!\!\!\!\!\!\!\!\!\!\! \{Definition \ref{reliable} and since $\ulcorner k_{as}^{-1}\urcorner$ = \{A, S\}\}&  \\
		& = & $\{A, S, B\}$~~~~~~~~~~~~~~~~~~~~~~~~(2.6)& \\
	\end{tabular}
}
$ $\\
4- Concordance with Theorem  LTWF:\\
$ $\\
From 2.1,  2.2, we have directly:\\
$$F'(N_a^i,r_{S_{A}^{2}}^+) \sqsupseteq \ulcorner N_a^i\urcorner \sqcap F'(N_a^i,R_{S_{A}^{2}}^-) ~~~~~~~~~~\mbox{(2.7)}$$
From 2.3 and  2.4 we have directly:\\
$$F'(X,r_{S_{A}^{2}}^+) \sqsupseteq \ulcorner X\urcorner \sqcap F'(X,R_{S_{A}^{2}}^-) ~~~~~~~~~~~~~~\mbox{(2.8)}$$
However, from 2.5 and  2.6, we declare that:\\
$$F'(Y,r_{S_{A}^{2}}^+) \not \sqsupseteq \ulcorner Y\urcorner \sqcap F'(Y,R_{S_{A}^{2}}^-) ~~~~~~~~~~~~~~~\mbox{(2.9)} $$ since we have no idea about the variable $Y$, thus, we have no idea about its value of security in the context (i.e. $\ulcorner Y \urcorner$ ).\\

From 2.9,  $S_{A}^{2}$ does not respect Theorem LTWF.  (II)\\

We abort our analysis here and we declare that the whole protocol does not respect Theorem LTWF.

\section{Discussion}
\normalsize

In our analysis, we have clearly established in 2.9 that the value of security of the variable $Y$ (which is an abstraction of the ticket $\{k_{ab}.A\}_{k_{bs}}$) goes down between the step 2 and 3 of the protocol. At this point, our witness function refused to certify the security of the protocol and flagged up a possible attack. In other words, the witness function pointed out that the variable $Y$  could fall in an intruder hands who may exploit it illegally, which is in tandem with the replay attack presented by Denning and Sacco in  \cite{Denning}. In fact, suppose an employee executing the first two steps of the protocol several times as a regular agent $A$ and collecting all the tickets $\{k_{ab}.A\}_{k_{bs}}$, as well as the corresponding session keys $k_{ab}$. If he is fired, as an external intruder $I$, he could still log on any server $B$ by playing the remaining steps of the protocol as follows:\\

\begin{tabular}{llllll}
	&$\langle3,$ & $I(A)$ & $\longrightarrow$ & $B:$ & $\{k_{ab}.A\}_{k_{bs}}\rangle$\\
	&$\langle4,$ & $B$ &$\longrightarrow$ & $I(A):$ & $\{N_b\}_{k_{ab}}\rangle$\\
	&$\langle5,$ & $I(A)$ &$\longrightarrow$ & $B:$ & $\{N_b - 1\}_{k_{ab}}\rangle$\\
\end{tabular}

$ $\\

To sum up, the Needham-Schroeder symmetric-key protocol neglected the fact that the  tickets $\{k_{ab}.A\}_{k_{bs}}$ is an important component and allowed it to circulate in clear in the step 3, which caused the described flaw to take place. 
This behavior is strictly forbidden in an analysis  using witness functions which never allows any message component to have a non increasing value of security even if it turns out that no actual flaw could be found. However, we can understand that back to the time when the protocol was designed, a few things were known about protocols, and protocol designers mostly thought that cryptography on its own was enough to ensure their security, which turned out to be incorrect over time.

\normalsize
\section{Comparison with related work }\label{sec5}

The analysis of cryptographic protocols remains a constraining task \cite{DBLP:conf/cade/Comon-Lundh08} despite all the progress made in the field due to the complexity and undecidability of the problem \cite{DBLP:journals/tocl/Comon-LundhCZ10}. Witness functions, using the general theorem or the little one, are beginning to gain ground and are proven to be a powerful means of detecting security breaches or designing correct protocols.  They have outperformed other analytical tools and methods such as interpretation functions \cite{Houmani1,Houmani3} in terms of efficiency and accuracy \cite{8123025}.  The fact that these functions allow a static analysis saves a lot of effort and time compared to other dynamic methods based on Model-Checking \cite{Basin2018} or on program logics \cite{DBLP:journals/pacmpl/SergeyWT18}, for example.  With that in mind, neither the witness functions nor any other verification method will  deliver its optimal protocol protection if other dimensions are neglected, especially security issues akin to the protocol implementation and encryption primitive weaknesses of the selected cryptographic system.

\section{Conclusion}\label{sec6}

In this paper, we have presented a detailed analysis of the Needham-Schroeder symmetric-key protocol using the little theorem of witness functions.  We have proven its ability to help detecting security vulnerabilities and inform about possible flaws. In future work, we intend to address the problem of  compose protocols \cite{DBLP:conf/esorics/HessMB18,BauerC016} as well as E-voting protocols \cite{CortierL18, DBLP:journals/ieeesp/RyanST15, CortierGT18} using these functions.



\bibliographystyle{ieeetr}


\bibliography{Ma_these}


\section*{Notations }
\begin{itemize}
\item[] $\ulcorner \alpha \urcorner$: initial value of security of an atomic message $\alpha$ in a context of verification.
\item[] $\sqcap$: minimum operator (the union in the security Lattice)
\item[] $\sqsupseteq$: greater than operator.
\item[] $\top$: highest value in the security Lattice (top).
\item[] $\bot$: lowest value in the security Lattice (bottom).
\item[] $r^+$: sent message in a generalized role.
\item[] $R^-$: received message in a generalized role.
\item[] $I$: intruder; $S$: honest server;  $A, B, ...$: principals.
\item[] $N_a$: nonce belonging to  $A$.
\item[] $k_{ab}$: key shared between  $A$ and $B$ (and $k_{ab}^{-1}$ its reverse form).
\item[] ${\mathcal{A}}{(m )}$: set of all atoms of a message $m$.
\item[] $\cal{A}_{\cal{G}}$: generalized role of an agent $A$.
\item[] $\cal{R}_{\cal{G}}(\textit{p})$: generalized roles of the protocol $p$.
\item[] $A$ $\longrightarrow B$  : $m$: $A$ sending a message $m$ to $B$.
\item[] $I(A)$: intruder impersonating (playing the role of) an agent $A$.
\end{itemize}

\end{document}